\documentstyle[sprocl,psfig]{article}

\input{epsf}

\def\np#1#2#3{           {\it Nucl. Phys. }{\bf #1} (19#2) #3}

\def\n.c.#1#2#3{         {\it Nuovo Cim. }{\bf #1} (19#2) #3}
\def\r.n.c.#1#2#3{       {\it Riv. del Nuovo Cim. }{\bf #1} (19#2) #3}

\relax

\newcommand{\GeV}{\,{\rm GeV}}

\newcommand{\beq}{\begin{equation}}
\newcommand{\eeq}{\end{equation}}
\newcommand{\beqa}{\begin{eqnarray}}
\newcommand{\eeqa}{\end{eqnarray}}
\newcommand{\ba}{\begin{array}}
\newcommand{\ea}{\end{array}}

\def\321{$SU(3)\times SU(2)\times U(1)$}

\def\n{\nu}

\def\r{\rho}

\def\be{\begin{equation}}
\def\ee{\end{equation}}
\def\bea{\begin{eqnarray}}
\def\eea{\end{eqnarray}}

\begin{document}

\title {R PARITY VIOLATION: CONSTRAINTS AND IMPLICATIONS}

\author{Anjan S. Joshipura}

\address{Dept. de Fisica Teorica,Univ. of
Valencia,\\461000,Burajssot, Valencia, Spain\footnote{
On leave from Physical Research Laboratory, Ahmedabad INDIA}.\\E-mail:anjan@prl.ernet.in}

\maketitle\abstracts{The constraints on trilinear
$R$ parity violating couplings $\lambda'_{ijk}$ following from 
({\em i}) the neutrino mass resulting due to  the induced
 vacuum expectation value for the sneutrino and ({\em ii}) the charm squark
interpretation for the HERA anomalous events are discussed in this talk.}

\section{Introduction}
The Baryon and the Lepton number symmetries enforced  by the gauge 
interactions and particle content in the standard model get broken when
it is extended to include supersymmetry. This violation 
is characterized in the minimal  supersymmetric standard model (MSSM) by

\begin{equation}
\label{wr}
W_{R} = -\tilde{\lambda}'_{ijk}~ L'_iQ'_{j} D'^{c}_{k}
-\tilde{\lambda}''_{ijk}~ U'^c_{i} D'^{c}_{j}
D'^c_{k}
-\tilde{\lambda}_{ijk}~ L'_{i} L'_{j}E'^c_{k}+\epsilon_i~ L'_iH_2~,
\end{equation}
where prime over the superfields indicates the weak basis and 
other notations are standard. The  couplings in (\ref{wr}) can be forbidden by 
imposing $R$ symmetry~\cite{r}. While the simultaneous presence of $\tilde{\lambda}''_{ijk}$ 
 and any of the other couplings is constrained severely by proton stability, 
the lepton number violating couplings by themselves are not
constrained as much. Their presence can lead to interesting signatures such as
 neutrino masses. We wish to discuss in this talk two topics related to the presence
 of the  the trilinear couplings ${\lambda}'_{ijk}$
namely, neutrino masses and possible anomaly seen in the 
$e^+p$ scattering at HERA \cite{hera}.

We shall specifically consider  the $\lambda'$-couplings related to the electron number 
violations as they are relevant for the description of HERA events. Moreover, they are 
also constrained 
more strongly than the others from the neutrino mass\cite{pap1,carlos}. 
We  first discuss these constraints
and their importance for the description of the HERA events and then specialize to the charm squark 
interpretation~\cite{pap2}. As we will discuss, this interpretation needs significantly large 
$\lambda'_{121}$ coupling in many models including the minimal supergravity based
scenario.

\section{Basis choice and definition of $\lambda'_{ijk}$}
In order to meaningfully constrain the trilinear coupling, it is sometimes assumed
 that
only a single coupling is non-zero at a time. While the physics implied
by these couplings is basis independent, the said assumption makes the 
constraints
on ${\lambda}'_{ijk}$ basis dependent since a non-zero $\lambda'$ in one basis 
may correspond to several non-zero $\lambda'$'s in the other. 

The relevant trilinear couplings
in eq.~(\ref{wr}) can be rewritten \cite{agashe} in  the quark mass basis as follows:
\begin{equation}
\label{basis1}
W_{R} = \lambda'_{ijk} (-\nu_id_{l}  K_{lj}+e_i u_j )d_k^c
\end{equation}
where $K$ denotes the standard Kobayashi Maskawa matrix. Even in the mass basis one could choose a different definition for the
trilinear couplings:
\begin{equation}
\label{lbar}
\bar{\lambda'}_{ijk}\equiv K_{jl}\lambda'_{ilk} 
\end{equation}
and rewrite (\ref{basis1}) as 
\begin{equation}
\label{basis2}
W_{R} =  \bar{\lambda'}_{ijk}(-\nu_id_{j}+e_i K^\dagger_{lj}u_l )d_k^c
\end{equation}
With the first choice, a single non-zero ${\lambda}'_{ijk}$ can lead to
 tree level flavour violations in the neutral
sector \cite{agashe} while this is not so if only one $\bar{\lambda}'_{ijk}\; (j\neq k)$
is non-zero. As an example of the basis dependence, let us note that the 
HERA results 
can be interpreted as
production of charm squark either by assuming only $\lambda'_{121}$ or 
$\bar{\lambda}'_{121}$ to be non-zero. The first coupling is constrained 
severely by the neutrino mass \cite{pap1} but the second is not. We shall return to this in section (4).
\section{Trilinear couplings and neutrino masses}
The presence of trilinear couplings generate neutrino masses in two different ways.
Firstly, eq.~(\ref{basis1}) directly leads to 1-loop diagrams generating neutrino masses.
This is a well-known contribution \cite{hs,probir}. But there is an additional
contribution \cite{pap1,carlos} which results from the following soft terms in the supersymmetry breaking part of
the scalar potential
\beq
\label{soft}
V_{soft} = - B_{\tilde{\nu}_{i}}~ \tilde{\nu}_{i} H_{2}^{0}  
+  m^{2} _{\nu_{i} H_{1}}~  \tilde{\nu}^{\star}_{i} H_{1}^{0} + c.c +\cdots~.
\end{equation}
Note that the $W_R$ in eq.(\ref{wr}) does not lead to the above soft
terms {\em at the GUT scale} in conventional supergravity based models
if $\epsilon_i$ are zero as assumed here. But terms in eq.(\ref{soft}) 
do get generated at the {\em weak scale} even in this case.
This fact becomes clear from the
 following
renormalization group equations \cite{pap1} satisfied by the soft parameters appearing 
in (\ref{soft}).
\begin{figure}[ht]
\vskip 1.0in
{\psfig{figure=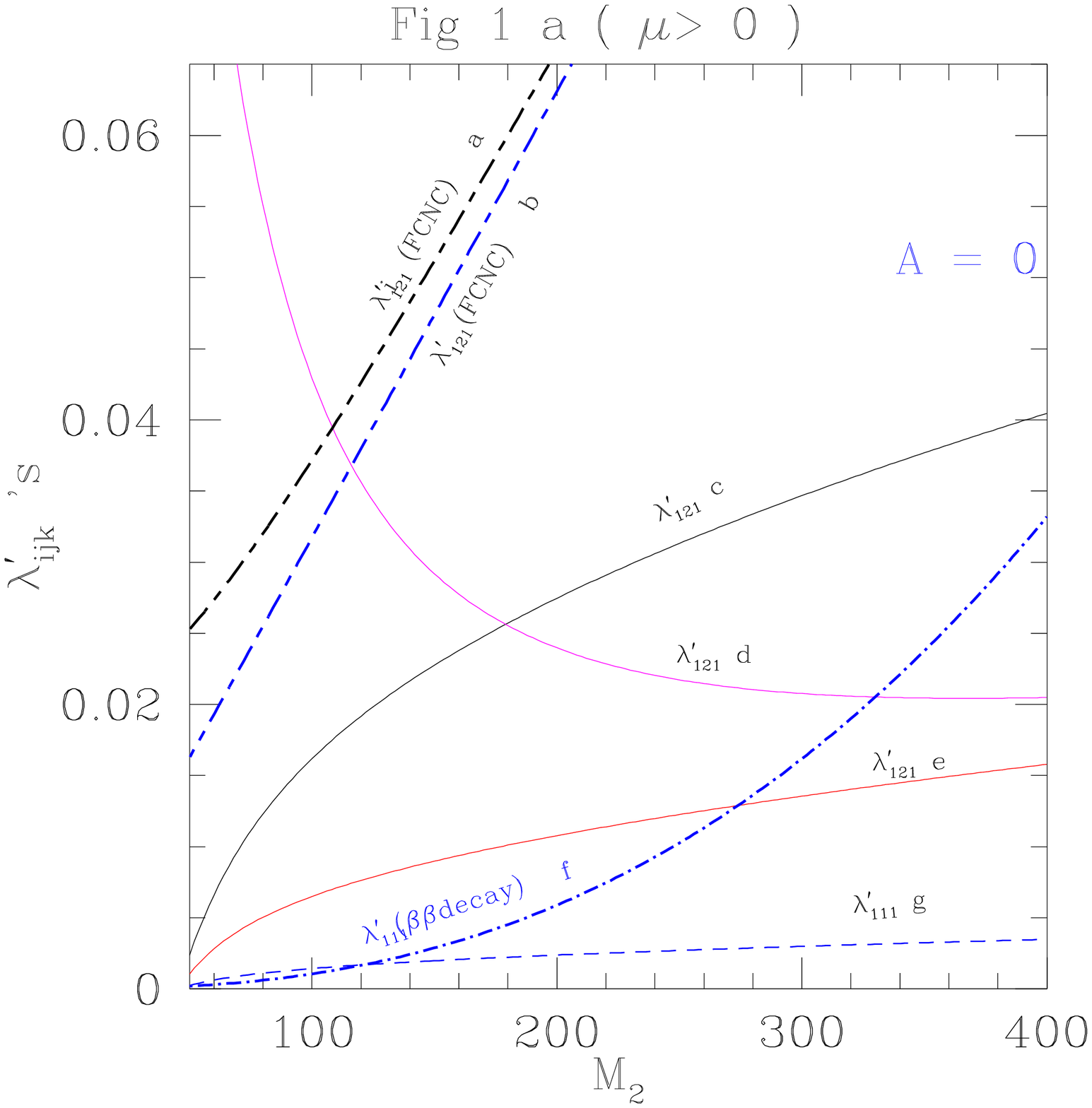,height=2.0in }} 
\vspace*{-2.00in}\hspace*{6.0cm}
{\psfig{figure=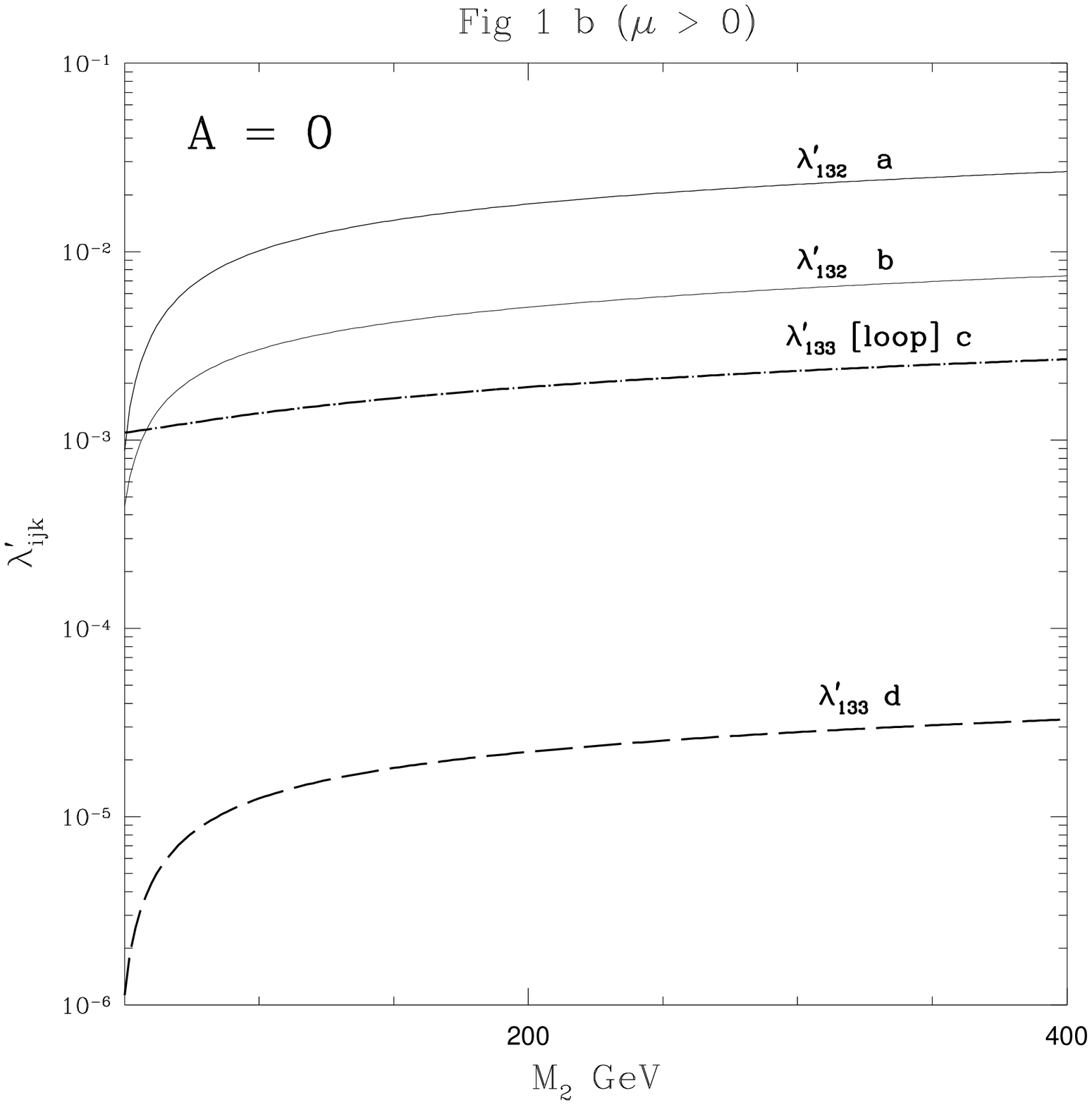,height=2.0in}}
{\bf Figure 1a}.~\sl{ FCNC constraints \cite{agashe} on $\lambda^{\prime}_{121}$ for a)~$m = 200$ GeV and 
b)~ $m = 50$ GeV for $tan\beta = 40$.
Neutrino mass constraints on $\lambda^{\prime}_{121}$ for 
c)~$m = 50$ GeV, $tan\beta = 15$; d)~$m = 200$ GeV ,$tan\beta = 40$ and	 
e)~$m = 50$GeV, $tan\beta = 40$.	 
f)~ $\lambda^{\prime}_{111}$ from neutrino less $\beta\beta$ decay \cite{dbd}
g)~$\lambda^{\prime}_{111}$ from neutrino mass constraints for $m = 50$ GeV 
and $tan\beta = 40$.}\\
{\bf Figure 1b}.~{\sl 
Neutrino mass constraints on $\lambda^{\prime}_{132}$ for a)~$tan\beta = 5$
and b)~$tan\beta = 25$; on $\lambda^{\prime}_{133}$ for $tan\beta = 5$ 
c)~considering only loop contributions and d)~loop as well as sneutrino VEV 
contributions . 
}

\end{figure}

\begin{eqnarray}
\label{rg}
\frac{dB_{\tilde{\nu}_i}}{dt}& =& -~ \frac{3}{2} B_{\tilde{\nu}_i}~\left(Y_{t}^{U} - \tilde{\alpha}
_{2} - \frac{1}{5} \tilde{\alpha}_{1}\right) - \frac{3 \mu}{16 \pi^{2}} 
{\lambda }_{ikk}^{\nu} h^{D}_{k} \left({A}^{\nu}_{ikk} + \frac{1}{2} B_{\mu} \right)~, \\
\frac{d m_{\nu_i H_{1}}^{2}}{dt}& = &-~ \frac{1}{2} m_{\nu_i H_{1}}^{2} \left(3 Y_{k}^{D} + 
Y_{k}^{E} \right)  
-\frac{3}{32 \pi^{2}} {\lambda }^{\nu}_{ikk}
h^{D}_{k} \left( m_{H_{1}}^{2} + m_{\tilde{\nu}}^{2} \right. \nonumber \\
& & \left. +2~ A_{ikk}^{\nu} A_{k}^{D} + 
2~  m_{k}^{Q{2}} +~ 2 ~m_{k}^{D^{c}2 } \right)~.
\end{eqnarray}
where $\lambda_{ikk}^\nu\equiv K_{lk} {\lambda}'_{ilk} $ and the terms on RHS are the standard soft supersymmetry breaking parameters.
It is clear that a non-zero $\lambda'_{ikk}$ generates non trivial $V_{soft}$
at the weak scale even when the parameters $B_{\tilde{\nu}_{i}},  m^{2} _{\nu_{i}H_1}$
 are zero at the GUT scale.
The $V_{soft}$ in eq.~(\ref{soft}) invariably induces the vacuum expectation value (vev) for the
sneutrino field and leads to a neutrino mass.
It turns out that due to additional logarithamic enhancement, this contribution to
the neutrino mass dominates over the loop induced contribution  in the supergravity 
based models. The constraints on $\lambda'_{1jk}$ following from this contribution are 
therefore stronger than from the loop induced contribution considered in the literature 
\cite{probir}.

We have adopted the minimal supergravity based scenario to explicitly derive these constraints.
Fig.~1a displays constraints on ${\lambda}'_{121},{\lambda}'_{111}$ for some values of the MSSM 
parameters and compares them with the existing constraints. It follows that constraints coming from 
the neutrino mass are quite strong and complimentary to the similar existing constraints.
Fig~1b shows similar constraints on parameters ${\lambda}'_{132},{\lambda}'_{133}$. More details can be 
found  in \cite{pap1}.
\section{Charm squark interpretation of the HERA events}
The interpretation of the HERA anomalies as due to resonant charm squark production 
requires \cite{pap2}
\begin{equation}
\label{heraeq}
\lambda'_{121}\sim \frac{0.025-0.034}{B^{1/2} }
\eeq
where $B$ is the branching ratio for the $R$ violating decay
$\tilde{c}_L\rightarrow e^+ d$. This equation implicitly depends upon the
parameters of the MSSM through $B$. These parameters must  be such  that the charm squark has the right mass
namely, around $200 \GeV$. Strictly speaking, charm squark mass can be treated as an
independent free parameter as has been done in recent studies \cite{pap2}. However
 this is not so
in a large class of models characterized by $m_{\tilde{c}_L}^2 (M_{GUT})>0$ and hence also
in the most popular minimal version of the supergravity based scenario.
Assuming unification of the gauge couplings 
and gaugino masses at the GUT scale $M_{GUT}=3 \times 10^{16}\GeV$, one has at a lower scale
$Q_0\sim 200 \GeV$
\beq
\label{mass}
m_{\tilde{c}_L}^2 (Q_0)\approx m_{\tilde{c}_L}^2 (M_{GUT})+
 8.83  M_2^2+ 1/2~ M_Z^2 ~\cos 2\beta~ (1-4/3 \sin^2\theta_W)^{1/2}
\eeq

\begin{figure}[ht]
\vskip 1.5cm \hspace*{3.5cm}
{\psfig{figure=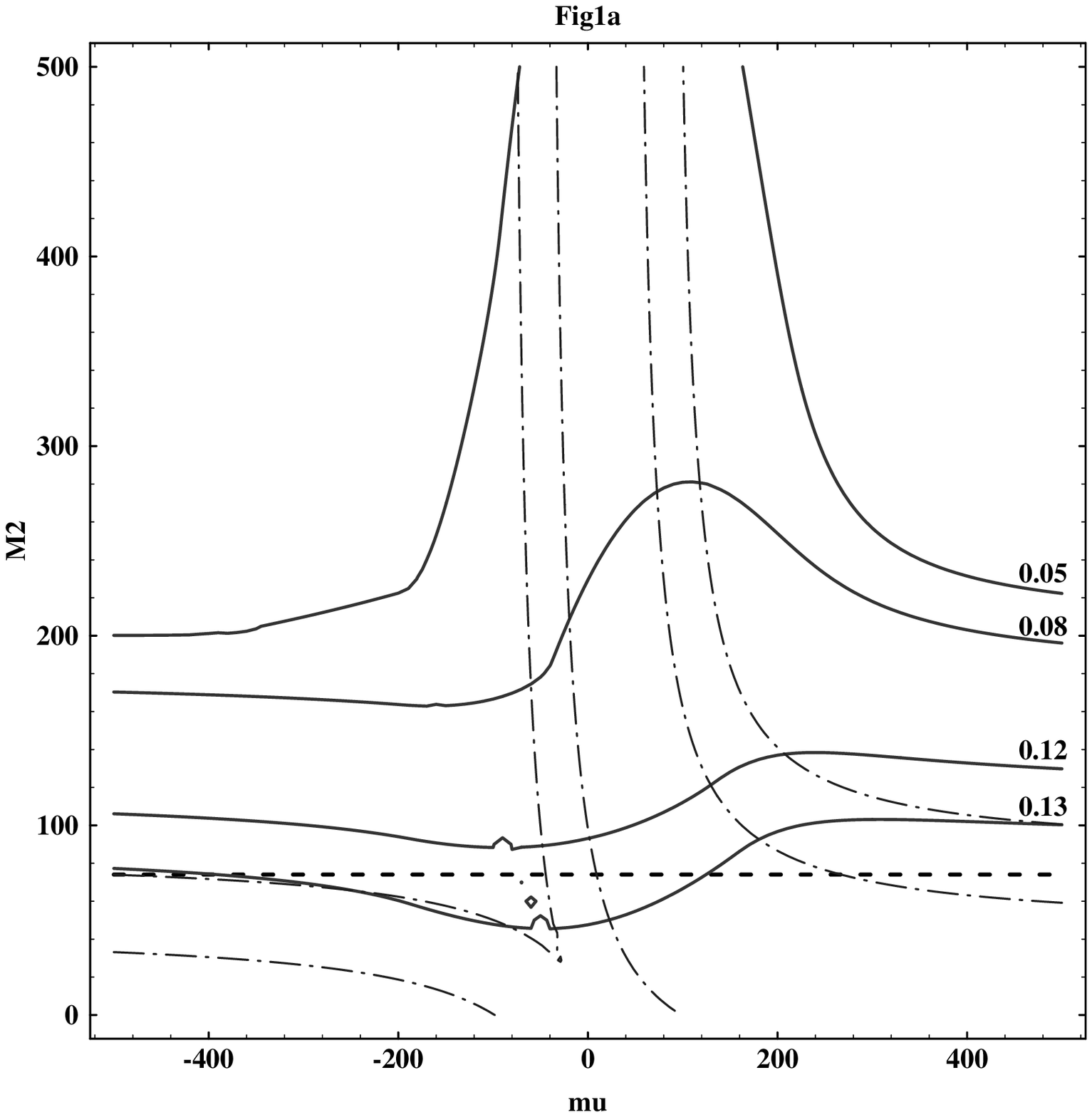,height=2.5in }}
\hspace*{-8.00cm}{\bf Figure 2}.~\sl{The contours (continuous lines ) of constant $\lambda'_{121}$  
obtained by imposing HERA constraint, eq.(8). The contours are for
values  0.05, 0.08, 0.12, and 0.13. 
The horizontal dashed line represents the bound on $M_2 $ 
coming from requiring $m_{\tilde{c}_L}=220 \GeV$. 
The vertical dash-dot lines represent the bounds on 
the chargino mass, the upper one for a mass of 85 \GeV and the
lower one for
a mass of 45 \GeV.  All the
above are computed for tan$\beta$ = 1. }

\end{figure}

Now $m_{\tilde{c}_L}^2 (M_{GUT})>0$ implies
\beq
\label{bound}
M_2\leq 74.04 \GeV \left({m_{\tilde{c_L}}\over 220 \GeV}\right )
\left ( 1- 0.06 \cos 2\beta \left({220 \GeV \over m_{\tilde{c_L}}}\right
)^2
\right)^{1/2}
\eeq

This bounded value for $M_2$ results in light chargino to which charm squark decays
dominantly reducing $B$ to a very small value \cite{pap2,dpm} and hence $\lambda'_{121}$ to a large
value through (\ref{heraeq}). This is quantitatively displayed in Fig.(2) where we plot contours of 
constant $\lambda'_{121}$ satisfying eq.(\ref{heraeq}). It is seen that the bound 
(\ref{bound}) does not 
allow $\lambda'_{121}$ to be small. The required large value of $\lambda'_{121}$
is severely constrained from the atomic parity violation \cite{dpa}, neutrino mass
\cite{pap1} and the decay
 $K\rightarrow \pi\nu\nu$ \cite{agashe}. One may try to avoid \cite{pap2} the last two bounds by choosing basis
given in (\ref{basis2}) and requiring that only $\bar{\lambda'}_{121}$ is non-zero.
But then one has  the following constraint coming from the neutrinoless 
double beta decay \cite{dbd}
in this case.
\beq
\label{dbd}
 K^{\dagger}_{12} \bar{\lambda'}_{121} \leq 2.2 \times 10^{-3}
\left({m_{\tilde{u}_L}\over 200
 \GeV}\right)^2 \left({ m_{\tilde{g}}\over 200
 \GeV}\right)^{1/2}
\eeq
This too does not allow $\bar{\lambda'}_{121}\sim O(0.1)$. One needs to allow
more than one non-zero $\lambda'_{121}$ and fine tune them \cite{pap2} to
 satisfy the neutrinoless double beta decay constraint.
\section{Summary}
We have underlined in this talk the phenomena of the generation of the sneutrino vev
\cite{pap1} and the resulting neutrino mass in the presence of trilinear $R$ violating couplings.
This additional contribution is shown to restrict the trilinear coupling much more 
strongly than
the corresponding loop contribution .  We have systematically derived these constraints.
We also discussed the charm squark interpretation of the HERA events. It was shown that such
interpretation requires large trilinear coupling in a wide class of models which include the minimal 
supergravity based model. Such a large coupling by itself is ruled out from other constraints but one may 
allow it by invoking new physics \cite{dpa} and postulating more than one non-zero couplings and fine tuning them.
\section*{Acknowledgments}
I thank my collaborators V. Ravindran and Sudhir Vempati for many interesting 
discussions . Thanks are also due to
D. Chaudhury, S. Raychaudhuri and J.W.F. Valle for helpful remarks.
This work was supported under DGICYT grant no. SAB-95-0627.

\end{document}